\documentclass[useAMS,usenatbib]{mn2e}
\usepackage{graphicx} 
\usepackage{color}
\usepackage{float}
\usepackage[justification=centering]{caption}
\usepackage{placeins}
\usepackage{multicol}
\usepackage{amsmath}
\usepackage{verbatim}
\usepackage{array}
\usepackage{soul}
\usepackage{booktabs}
\usepackage{threeparttable}
\usepackage{amssymb}
\usepackage{pifont}
\usepackage{array,multirow}
\def\be{\begin{equation}} 
\def\ee{\end{equation}}

\def\msun{{\Msun}}

\def\gsim{\lower.5ex\hbox{\gtsima}} 
\def\lsim{\lower.5ex\hbox{\ltsima}} \def\gtsima{$\; \buildrel > \over 
\sim \;$} \def\ltsima{$\; \buildrel < \over \sim \;$} \def\prosima{$\; 
\buildrel \propto \over \sim \;$} \def\gsim{\lower.5ex\hbox{\gtsima}} 
\def\lsim{\lower.5ex\hbox{\ltsima}} 
\def\simgt{\lower.5ex\hbox{\gtsima}} 
\def\simlt{\lower.5ex\hbox{\ltsima}} 
\def\simpr{\lower.5ex\hbox{\prosima}} \def\la{\lsim} \def\ga{\gsim} 
  
 \def\gtsima{$\; \buildrel > \over \sim \;$} 
\def\ltsima{$\; \buildrel < \over \sim \;$} 
\def\gsim{\lower.5ex\hbox{\gtsima}} 
\def\lsim{\lower.5ex\hbox{\ltsima}} 
\def\simgt{\lower.5ex\hbox{\gtsima}} 
\def\simlt{\lower.5ex\hbox{\ltsima}} 
\def\simpr{\lower.5ex\hbox{\prosima}} 
\def\la{\lsim} 
\def\ga{\gsim}

\def\msun{\,{\rm \Msun}}

\def\E3{{\cal E}_{\rm g}^{III}}

\def\Msun{\rm M_\odot}

\def\Msun{\rm M_\odot}
\def\Lsun{\rm L_\odot}
\def\myr{\rm Myr }

\def\M*{M_*}
\def\Z*{Z_*}
\def\L*{L_*}

\title[AGN feedback in high-z galaxies]{The impact of black hole feedback on the UV luminosity and stellar mass assembly of high-redshift galaxies}
\author[Piana et al.]{Olmo Piana$^{1}$\thanks{piana@astro.rug.nl}, Pratika Dayal$^1$, Tirthankar Roy Choudhury$^2$\\ 
\\
$^{1}$ Kapteyn Astronomical Institute, University of Groningen, P.O. Box 800, 9700 AV Groningen, The Netherlands \\
$^{2}$National Centre for Radio Astrophysics, Tata Institute of Fundamental Research, Pune 411007, India\\
}

\begin{document} 
 
\date{} 

\maketitle

\begin{abstract}
We employ the Delphi semi-analytical model to study the impact of black hole growth on high-redshift galaxies, both in terms of the observed UV luminosity and of the star formation rate. To do this, firstly, we assess the contribution of AGN to the total galaxy UV luminosity as a function of stellar mass and redshift. We find that for $M_\mathrm{UV} \lsim -24$ mag and $z \approx 5 - 6$ the galaxies for which the black hole UV luminosity outshines the stellar UV emission become the majority, and we estimate their duty cycle. Secondly, we study the evolution of the AGN and stellar luminosity functions (LFs), finding that it is driven both by changes in their characteristic luminosities (i.e.~evolution of the intrinsic brightness of galaxies) and in their normalizations (i.e.~evolution of the number densities of galaxies), depending on the luminosity range considered. Finally, we follow the mass assembly history for three different halo mass bins, finding that the magnitude of AGN-driven outflows depends on the host halo mass. We show that AGN feedback is most effective when the energy emitted by the accreting black hole is approximately $1\%$ of the halo binding energy, and that this condition is met in galaxies in halos with $M_\mathrm{h} \sim 10^{11.75} \msun$ at $z=4$. In such cases, AGN feedback can drive outflows that are up to 100 times more energetic than SN-driven outflows, and the star formation rate is a factor of three lower than for galaxies of the same mass without black hole activity.
\end{abstract}


\section{Introduction}
The UV luminosity function (UVLF) is a key observational probe of galaxy formation at high redshifts, thanks to important ground-based and space-based facilities such as the HST (Hubble Space Telescope), the VLT (Very Large Telescope), Alma and Subaru telescopes. The stellar UV luminosity can be used as a proxy for the underlying star formation activity \citep{ouchi2005, lee2009, mclure2009, stark2009, labbe2010a, gonzalez2010, mclure2013, finkelstein2015, bouwens2015, livermore2017, ishigaki2018, ono2018} and it also seems to correlate with other physical properties such as the size of the host galaxy \citep{kawamata2018}. Several works have put a lot of effort into studying the galaxy UVLF at different epochs, characterising its shape in the redshift range $z = 4 - 8$ \citep{castellano2010, mclure2010, oesch2014, atek2015, bouwens2015, bowler2015, mcleod2016, livermore2017, bouwens2017, ishigaki2018}. Nevertheless, as pointed out by \cite{ono2018} and \cite{harikane2021}, its bright end still suffers from contamination coming from misidentified AGN and strongly-lensed sources, as well as from inaccuracies given by photometric redshift determination and low statistics. Alongside the galaxy UVLF, the high-redshift AGN UVLF has been studied to gain an understanding of the evolution of supermassive black holes and of the contribution of quasars to reionisation with contrasting results \citep[][]{madau2004, giallongo2015, grazian2018, dayal2019, trebitsch2018, trebitsch2020, kim2020}, if we assume that the AGN emission tracks the accretion rate of the central black holes. The study of the AGN UVLF, though, is complicated by the obscuration effects of the torus surrounding the central black hole, which depend on its orientation with respect to the line of sight \citep{antonucci1985, urry1995}, and this might result in incomplete AGN samples. For this reason, and because of low statistics, constraints on the AGN UVLF are looser, and it is still unclear whether the collected data are better fitted by a Schechter function or by a double power-law \citep{willott2010, mcgreer2013, kashikawa2015, jiang2016, yang2016, manti2017, parsa2018, kulkarni2019}. Being able to disentangle the UV contributions from the AGN and stellar components to the total UV luminosity would help in setting more precise constraints on the bright end of the galaxy and AGN UVLF \citep{ono2018}. 

Observations of luminous accreting supermassive black holes at $z > 7$ \citep{mortlock2011, banados2018} showed that these objects are able to emit enormous amounts of energy, up to $10^{46}\ \mathrm{erg\ s^{-1}}$, supposedly exceeding the power emitted by Supernovae (SN) and leaving an imprint on their assembly histories. It has been suggested that this activity might be the cause for the correlations between the physical properties of the black holes and those of the host galaxies observed at low redshift \citep[see][for a review]{kormendy2013}. The absolute magnitude of the bulges is found to be proportional to the logarithm of the mass of the black holes already in \cite{kormendy1995}. Later works confirmed this finding and unveiled even tighter correlations between black hole masses and the mass, luminosity and velocity dispersion of the bulges \citep{magorrian1998, ferrarese2000, gebhardt2000, tremaine2002, haring2004, gultekin2009, kormendy2013, mcconnell2013, savorgnan2015}, finding mass ratios $M_{\mathrm{bh}}/M_{\mathrm{bulge}} \approx 0.002-0.006$ with a scatter of $\approx 0.25-0.3$ dex. Nevertheless, the evolution of these relations remains an open question at $z>4$. It has also been shown that numerical simulations need to incorporate AGN feedback - able to quench star formation - in order to explain the statistical properties of massive elliptical galaxies and the offset between the high-mass end of the halo mass function and that of the galaxy mass function \citep{silk-rees1998, dimatteo2005, croton2006, bower2006, sijacki2007, vogelsberger2014, schaye2015}. The implementation of negative AGN feedback (i.e.~able to suppress star formation) also appears to yield galaxy scaling relations which are in better agreement with observations \citep{dubois2013}. It is important to note, though, that AGN feedback is not the only viable quenching mechanism, and simulations that fail to account for this might overestimate the impact of AGN feedback. For instance, other simulations \citep[e.g.][]{dekel2006, bower2017} show that the transition of galaxies at $z \lsim 3$ from cold to hot accretion mode might also play an important role in galaxy quenching, by hindering the accretion of filaments of cold gas that would provide fuel for star formation for stellar masses of $M_* \gtrsim 3 \times 10^{10} \msun$. In addition, numerical simulations and semi-analytic models often struggle in including environmental effects such as strangulation, ram-pressure stripping, tidal stripping, and harassment, which have been shown to be responsible for emptying galaxies of their gas and quenching the star formation activity at lower stellar masses and redshift \citep{hirschmann2014}. For this reason, it is still unclear what role these mechanisms play at high redshifts. 

Observationally, widespread detection of AGN-driven outflows in active galaxies at $z \lsim 3$ \citep{fischer2010, feruglio2010, greene2011, aalto2012, harrison2012} seemed to support the hypothesis that AGN feedback might remove molecular gas from the disk of the host galaxies and quench the star formation activity. On the other hand, outflows at higher redshift have hardly been detected \citep{maiolino2012, cicone2015}, though stacking spectral analysis performed on AGN samples at $z \sim 6$ show signatures of extended outflows \citep{gallerani2018, stanley2019}. Some studies have tried to establish correlations between the presence of these outflows and the star formation rates of the galaxies, but the results remain inconclusive so far: thanks to integral field units techniques, local spatial anti-correlation between AGN-driven outflows and the star formation rate within the galaxy has been found \citep[e.g.][]{canodiaz2012}, but whether these outflows are able to quench star formation on a galactic scale is still unclear. At the same time, signs of an enhancement of star formation activity due to increased gas pressure in outflows have also been observed \citep{maiolino2017}, and in some cases the simultaneous presence of negative and positive feedback within the same galaxy was revealed \citep{cresci2015}, confirming the prediction of \cite{silk2013}. Unfortunately, these observations often come with several caveats. The determination of reliable star formation rates depends on the ability to accurately model dust extinction, or on the availability of high-quality far-infrared observations. In addition, as hinted earlier, building statistically significant samples of AGN and star-forming galaxies of similar stellar masses to compare with each other has proven to be difficult, given that precise stellar mass measurements require accurate dust and stellar population modelling, besides being able to accurately distinguish between the AGN and the stellar emissions. 

In this paper we follow-up on the work presented in \cite{piana2021}, where we have shown the assembly history of black holes during the first billion years of the Universe. While this first work focused on the description of the statistical properties of the high-redshift black hole population, here we want to characterise the connection between the black hole and the host galaxy. The goal of this paper is first to assess the importance of the AGN contribution to the total UV luminosity of galaxies as a function of their total luminosity (Section~\ref{UV LF_dis}), in order to provide a tool to disentangle the stellar and the AGN UV luminosity functions, and secondly to study at what galaxy mass range AGN feedback are most efficient in driving outflows, strongly affecting the star formation history of the host galaxy (Section~\ref{bh_impact}).

The cosmological parameters used in this work are $\Omega_{\rm m }, \Omega_{\Lambda}, \Omega_{\rm b}, h, n_s, \sigma_8 = 0.3089, 0.6911, 0.049, 0.67, 0.96, 0.81$ \citep{planck2015}. We quote all quantities in comoving units, unless stated otherwise, and express all magnitudes in the standard AB system \citep{oke1983}.


\section{The theoretical model}
\label{sec_model}

\begin{table*}\begin{center}
\begin{threeparttable}
\setlength{\extrarowheight}{3pt}
\caption{The model parameters shown in column 1 are described column 2. Columns 3 and 4 show the free parameter values used for these  when applying AGN visibility corrections using the results of \protect\cite{ueda2014} and \protect\cite{merloni2014}, respectively.}
\begin{tabular}{ccccc}
\hline
\hline
Parameter & & Description & $f_\mathrm{unobs}$ [Ueda+ 2014] & $f_\mathrm{unobs}$ [Merloni+ 2014]\\
\hline
$\epsilon_\mathrm{r}$ & & radiative efficiency of black hole accretion & 0.1 & 0.1\\
$f_*$ & & star formation efficiency threshold & 0.02 & 0.02\\
$f_*^\mathrm{w}$ & & fraction of SN energy that couples to the gas & 0.1 & 0.1\\
$f_\mathrm{bh}^\mathrm{w}$ & & fraction of AGN energy that couples to the gas & 0.003 & 0.0035\\
$f_\mathrm{bh}^\mathrm{ac}$ & & fraction of available gas mass that black holes can accrete & $5.5 \times 10^{-4}$ & $10^{-4}$\\
$f_\mathrm{Edd} (M_\mathrm{h} < M_\mathrm{h}^\mathrm{crit})$ & & black hole accretion rate in fraction of Eddington & $7.5 \times 10^{-5}$ & $7.5 \times 10^{-5}$\\
$f_\mathrm{Edd} (M_\mathrm{h} > M_\mathrm{h}^\mathrm{crit})$ & & black hole accretion rate in fraction of Eddington & 1 & 1\\
$\alpha$ & & LW background threshold for DCBH formation (in units of $J_{21}$) & 30 & 30\\
\noalign{\smallskip}
\hline
\hline
\end{tabular}

\label{table_models}
\end{threeparttable}
\end{center}\end{table*}

In this work we use the semi-analytic code \textit{Delphi} (\textbf{D}ark matter and the \textbf{e}mergence of ga\textbf{l}axies in the e\textbf{p}oc\textbf{h} of re\textbf{i}onization) to study the role of star formation and black holes and their associated feedback in determining both the UV luminosities and physical properties of early galaxies \citep{dayal2014, dayal2019}. We briefly describe the model here and interested readers are referred to our earlier papers \citep{dayal2019, piana2021} for full details. Our model is based on semi-analytic dark matter halo merger trees which track the accretion and merging history of 550 $z=4$ dark matter halos with (equally separated) mass values between $\log (M_\mathrm{h}/\msun) = 8-13.5\ \Msun$. We use a binary merger tree algorithm \citep{parkinson2008} to reconstruct merger trees up to $z=20$ in timesteps of 20 Myr. Each halo at $z=4$ is assigned a number density according to the Sheth-Tormen halo mass function (HMF) at $z=4$, and this number density is propagated throughout all of its merger tree so that the Sheth-Tormen HMF is matched at each redshift. At each time step, dark matter halos can grow both via smooth accretion from the intergalactic medium (IGM) and mergers with other halos. Similarly, the gas mass contained in the galaxy is replenished both through smooth accretion of gas from the IGM and through mergers; for the former, we make the reasonable assumption that smooth accretion of dark matter is accompanied by accretion of a cosmological fraction of gas mass. The starting leaves of the merger trees are seeded with black holes: in the {\it fiducial} model, these seeds are $10^{3-4}\msun$ direct-collapse black holes (DCBHs) if the virial temperature of the halo is $\gsim 10^4$ K and the Lyman-Werner (LW) background impinging on the halo is $30 J_{21}$, where $J_{21}$ is the LW background expressed in units of $10^{-21} {\rm erg\, s^{-1}\, Hz^{-1} \, cm^{-2} \, sr^{-1}}$ \citep{sugimura2014}. Starting leaves at $z>13$ not meeting these criterion are instead seeded with a $150\msun$ stellar black hole (SBH). 

The initial gas mass can form stars with an effective efficiency $f^{\mathrm{eff}}_*$. This is defined as the minimum value between the star formation efficiency whose corresponding SN-II feedback is enough to expel the rest of the gas from the host galaxy and an upper threshold value ($f_* = 2\%$). A fraction of the gas mass left after star formation and SNII gas ejection can be accreted onto the black hole. This accreted mass is given by
\begin{equation}
M_{\mathrm{bh}}^{\mathrm{ac}}(z) = \min\left[f_{\mathrm{Edd}}M_{\mathrm{Edd}}(z),\ f_{\mathrm{bh}}^{\mathrm{ac}}M_{\mathrm{g}}^{\mathrm{sn}}(z) \right],
\label{eq_min}
\end{equation}
where $M_{\mathrm{Edd}} = \dot{M}_{\mathrm{Edd}} \Delta t$ ($\Delta t = 20\ \myr$) is the Eddington mass, $f_\mathrm{Edd}$ is the Eddington fraction, $f_\mathrm{bh}^\mathrm{ac}$ represents the maximum fraction of the total gas mass left after star formation and SNII feedback ($M_{g}^\mathrm{sn}$) that can be accreted onto the black hole, and $\epsilon_\mathrm{r} = 0.1$ is the radiative efficiency of the black hole. The value of $f_\mathrm{Edd}$ depends on the halo potential, given that black hole growth is stunted in halos below a characteristic mass  $M_{\mathrm{h}}^{\mathrm{crit}}(z) = 10^{11.25}\msun[\Omega_\mathrm{m}(1+z)^3 + \Omega_\lambda]^{0.125}$ \citep[see][]{bower2017, habouzit2017}. In our model, $f_{\mathrm{Edd}}$ is a step function such that  
\begin{equation}
f_{\mathrm{Edd}} = \begin{cases} 
      7.5 \times 10^{-5} & M_\mathrm{h} < M_\mathrm{h}^{\mathrm{crit}} \\
      1 & M_\mathrm{h} \geq M_\mathrm{h}^{\mathrm{crit}} \\
   \end{cases}
   \label{eq_fedd}
\end{equation}
The black hole mass accretion rate at any time-step is then $\dot{M}_{\mathrm{bh}} = M_{\mathrm{bh}}^{\mathrm{ac}}(z)/\Delta t$. A fraction $f_{\mathrm{bh}}^\mathrm{w}$ of the total energy emitted by the accreting black hole couples to the gas and drives outflows. The remaining gas mass, $M^{\mathrm{gf}}_{\mathrm{bh}}$, represents the final gas mass at the end of the time step. Our model parameters ($f_*^\mathrm{w}$, $f_\mathrm{bh}^\mathrm{w}$, $f_\mathrm{bh}^\mathrm{ac}$ and $f_\mathrm{Edd}$) are tuned to simultaneously reproduce the key observables for both star forming galaxies (such as the evolving ultra-violet LF, stellar mass function, star formation rate density) and AGN (the AGN UVLF and the black hole mass function). While we explicitly include visibility corrections for black holes as described in what follows, the star formation efficiency has been tuned to reproduce the {\it observed} UVLF of Lyman-break galaxies (LBGs) and should therefore be thought of as the dust-attenuated (observed) value. For this reason, we do not include any dust correction for star formation in this work. We summarise the values of the key model parameters in Table \ref{table_models}. 

Each new stellar population is assumed to form following a Salpeter initial mass function \citep{salpeter1955} in the mass range $0.1 - 100 \msun$, with a fixed metallicity $Z = 0.05\ Z_\odot$ and an age of $2\ \myr$. Its time-dependent UV luminosity $L^\mathrm{UV}_*$ is computed using \textit{STARBURST99} \citep{leitherer1999}, and decays with increasing age of the stellar population. Hence, the total stellar luminosity of a galaxy at a specific time step is the sum of the contribution coming from the new star forming event plus that from the star forming events occurred within each of the progenitors during the previous time step, such that $L_\mathrm{*, tot}^{\mathrm{UV}}(z) = L_{*}^{\mathrm{UV}}(z) + \sum L_\mathrm{*, tot}^{\mathrm{UV}}(z+\Delta z)$, with the sum running over all the progenitors of the galaxy. With respect to the black holes, instead, we can write their bolometric luminosity as
\begin{equation}
L_{\mathrm{bh}} = \frac{\epsilon_\mathrm{r} \left(1 - f_{\mathrm{bh}}^\mathrm{w} \right) \dot{M}_{\mathrm{bh}}c^2}{\Delta t}.
\end{equation} 
We first convert this bolometric luminosity into the B-band luminosity as 
\begin{equation}
\log \left(\frac{L_{\mathrm{bh}}}{\nu_\mathrm{B} L_{\mathrm{bh}}^{\nu_\mathrm{B}}}\right) = 0.80 - 0.067 \Lambda + 0.017 \Lambda^2 - 0.0023 \Lambda^3, 
\end{equation}
where $\Lambda = \log \left(L_{\mathrm{bh}}/\Lsun \right) - 12$ and $L_{\mathrm{bh}}^{\nu_\mathrm{B}}$ is expressed in solar luminosities \citep{marconi2004}. Assuming an AGN spectral slope of $L_{\mathrm{bh}}^\nu \propto \nu^{-0.44}$, we can convert the B-band luminosity to UV luminosity at $1375\ \mathrm{\AA}$ ($L_{\mathrm{bh}}^{\mathrm{UV}}$). The total UV luminosity of the galaxy is then defined as
\begin{equation}
L_{\mathrm{tot}}^{\mathrm{UV}}(z) = L_\mathrm{*, tot}^{\mathrm{UV}}(z) + L_{\mathrm{bh}}^{\mathrm{UV}}(z).
\end{equation}
Note that in this formulation we are assuming that the black hole luminosity decays instantaneously.


\begin{figure}
\includegraphics[width=0.5\textwidth]{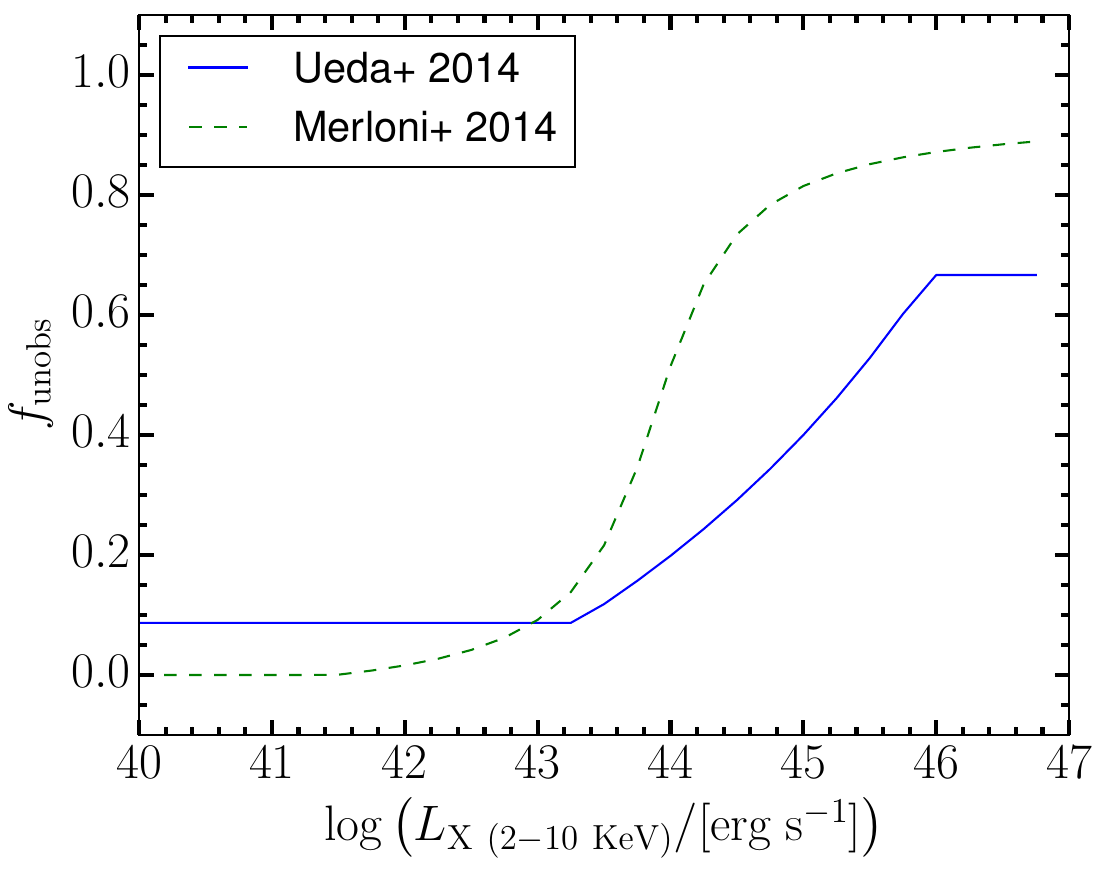}\\
\caption{The redshift-independent unobscured AGN fractions as a function of the X-ray luminosity inferred by \citep[][solid line]{ueda2014} and \citep[][dashed line]{merloni2014}.} 
\label{funobs}
\end{figure}

\begin{figure*}
\scalebox{1}[1]{\includegraphics[width=\textwidth]{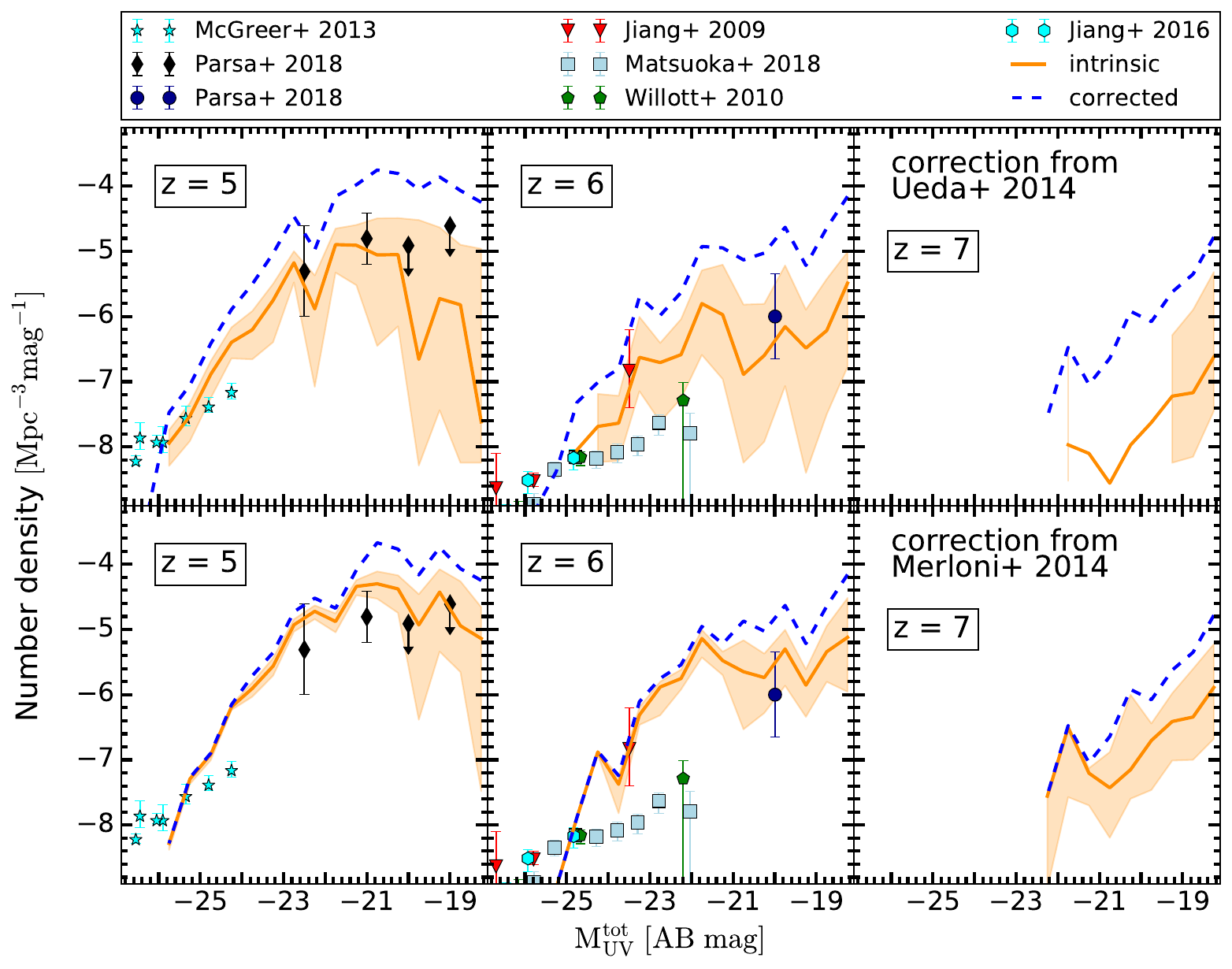}}\\
\caption{We plot here the intrinsic (dashed lines) and corrected (solid lines) AGN UV luminosity functions at $z = 5, 6$ and 7, using both \citep[][upper panels]{ueda2014} and \citep[][lower panels]{merloni2014} corrections. The extent of the shaded areas for the corrected UVLF represents the difference between the 84th and 16th percentile (i.e.~the 1-sigma error) of the distribution in each bin for 1000 realisations of the two black hole obscuration models considered. Observational constraints are also shown \citep{jiang2009, willott2010, mcgreer2013, jiang2016, parsa2018, matsuoka2018}.} 
\label{agnuvlf}
\end{figure*}

A certain fraction of the total AGN population is obscured by the torus and dust surrounding the central engine. While observations point to the fraction of obscured AGN evolving with the AGN luminosity, these are limited to $z \lsim 2$ \citep[e.g.][]{ueda2003, lafranca2005, hasinger2008, iwasawa2012, ueda2014}. In this work, we use the obscured fractions computed in the works by \cite{ueda2014} and \cite{merloni2014}, which are defined as functions of the X-ray luminosity of the black holes $L_X$ computed across the energy range $2-10$ keV. Defining unobscured AGN to be those with a column density $\log(N_H) < 22$, \citep{ueda2014} find an unobscured fraction 
\begin{equation}    
f_{\mathrm{unabs}} = \frac{1 - \psi}{1 + \psi},
\end{equation}
where $\psi = 0.43 \left[1 + \min(z, 2) \right]^{0.48} - 0.24 \left(\log L_X-43.75 \right)$. This function saturates at $0.008$ at the low-luminosity end and at $0.73$ at the high-luminosity end. On the other hand, \cite{merloni2014} find 
\begin{equation}
f_{\mathrm{unobs}} = 1 - 0.56 + \frac{1}{\pi}\arctan \left(\frac{43.89 - \log L_X}{0.46} \right).
\end{equation}
These unobscured fractions are shown in Fig.~\ref{funobs}. As seen, the results from \cite{ueda2014} yield a higher (lower) fraction of unobscured AGN as compared \cite{merloni2014} corrections below (above) $L_X \sim 10^{43}\ \mathrm{erg\ s^{-1}}$. This necessitates a slight change in the AGN-associated free parameters between these models, as shown in Table~\ref{table_models}, to match to the observed AGN UV LF at $z \sim 5-7$. To implement these results in our model, we proceed as follows: in any given bin, we randomly select a fraction $f_{\mathrm{unobs}}(L_X)$ of AGN, for which we assume that all the luminosity from the central black hole reaches the observer; the rest of the AGN are assumed to be completely obscured, with $L^\mathrm{bh}_\mathrm{UV} = 0$. This implies that the total UV Luminosity is $L^{\mathrm{UV}}_{\mathrm{tot}} = L^{\mathrm{UV}}_* + L^{\mathrm{UV}}_{\mathrm{bh}}$ for unobscured AGN and  $L^{\mathrm{UV}}_{\mathrm{tot}} = L^{\mathrm{UV}}_* $ for obscured AGN \citep[see also][]{volonteri2017, ricci2017, dayal2019}. This process introduces some fluctuations in the estimation of the luminosity function since black holes falling in the same luminosity bin do not necessarily share exactly the same number density (especially at low masses). To solve this issue, we perform 1000 realisations of the sampling procedure and compute the median AGN number density in each luminosity bin. We plot the resulting corrected AGN UV luminosity function in  Fig.~\ref{agnuvlf} (solid orange line), together with the intrinsic AGN UVLF (dashed blue line). We can see that our corrected AGN UV LF reproduces the bulk of the black hole population at $z = 5$ and 6, down to $M_\mathrm{UV} = -26$ mag, and the implemented obscuration effects are needed particularly to match the observations at $M_\mathrm{UV} \gtrsim -23$ mag.

\begin{figure*}
\includegraphics[width=\textwidth]{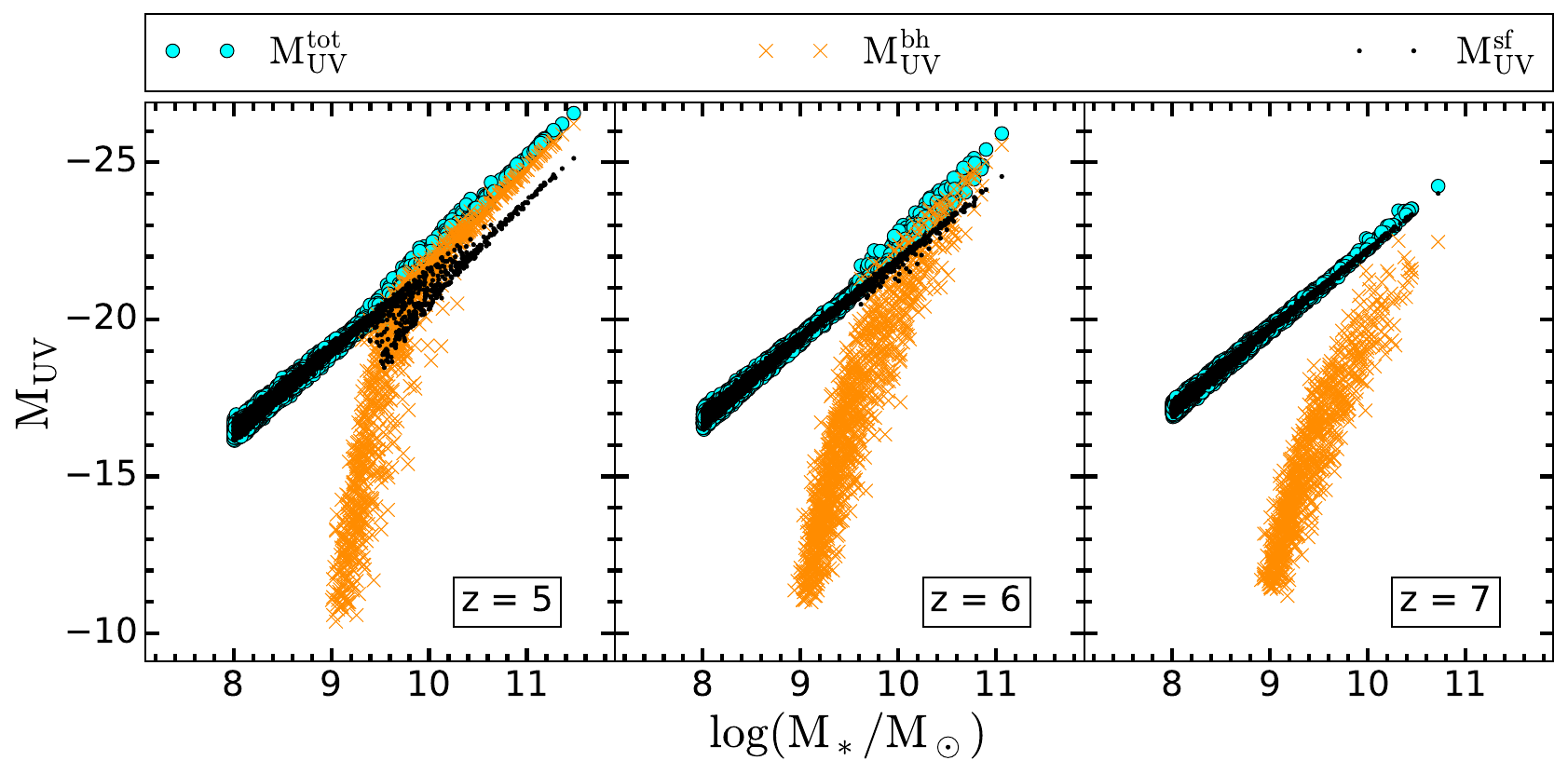}\\
\caption{The $M_\mathrm{UV}-M_*$ relation at $z = 5, 6$ and 7. For each galaxy, we show the {\it intrinsic} UV magnitudes summing over star formation and black hole accretion (cyan filled circles), star formation only (black dots) and black hole accretion only (orange crosses). Notice that the black dots overlap with the cyan circles at high redshift and low stellar masses, where the contribution from black hole activity to the UV luminosity is negligible.}
\label{uvmag_smass}
\end{figure*}

In order to test the robustness of the results of our model, we compare our {\it fiducial} model ({\it ins1}), in which black holes are assumed to merge at the same time as their parent halos, to the ({\it tdf4}) scenario, in which we implement additional features: (i) a delay between galaxy mergers and the corresponding black hole merger due to dynamical friction during the inspiraling phase; (ii) a maximal reionization feedback that photoevaporates the gas mass in halos with rotational velocities $v_\mathrm{c} < 40\ \mathrm{km\ s^{-1}}$; (iii) a higher threshold of Lyman-Werner background for DCBH seed formation, with $\alpha = 300 J_{21}$. Specifically, in the second scenario, the ``merging" timescale is calculated as \citep{lacey1993}
\begin{equation}
\tau = f_\mathrm{df} \Theta_\mathrm{orbit} \tau_\mathrm{dyn} \frac{M_\mathrm{host}}{M_\mathrm{sat}} \frac{0.3722}{\ln(M_\mathrm{host}/M_\mathrm{sat})},
\end{equation}
where $M_\mathrm{host}$ is the mass of the host including all the satellites, $M_\mathrm{sat}$ is the mass of the merging satellite, $\tau_\mathrm{dyn}$ represents the dynamical timescale and $f_\mathrm{df}$ represents the efficiency of tidal stripping; $f_\mathrm{df}>1$ if tidal stripping is very efficient. For this work, we use $f_\mathrm{df}=1$. Further, 
\begin{equation}
\Theta_{orbit} = \bigg(\frac{J}{J_\mathrm{c}}\bigg)^{0.78} \bigg(\frac{r_\mathrm{c}}{R_\mathrm{vir}}\bigg)^2,
\end{equation}
where $J$ is the satellite's specific angular momentum and $J_\mathrm{c}$ that of a satellite carrying the same energy and orbiting on a circular orbit. The last term represents the ratio between the circular radius (the radius of a circular orbit with the same energy) and the virial radius of the host. $\Theta_\mathrm{orbit}$ is well modelled by a log-normal distribution such that ${\rm log} (\Theta_\mathrm{orbit}) = -0.14 \pm 0.26$ \citep{cole2000}; we randomly sample values from this distribution for each merger. Finally, the dynamical timescale can be calculated as $\tau_\mathrm{dyn} = \pi R_\mathrm{vir}(z) V_\mathrm{vir}(z)^{-1} = 0.1 \pi t_\mathrm{H}(z)$ where $R_\mathrm{vir}$ and $V_\mathrm{vir}$ are the virial radius and velocity at $z$. We also make the limiting assumption that satellite galaxies, waiting to merge, neither form stars nor have any BH accretion of gas. The results of such modified scenario are detailed in Sec.~\ref{bh_impact}. The consequences of the {\it tdf4} assumptions mostly affect low-mass galaxies: reionization feedback suppresses the gas masses of the lowest-mass galaxies. This means that, when merging, these galaxies will not bring in any gas mass, delaying the buildup of the gas, stellar and black holes components at early times. At the same time, the delay in mergers induced by dynamical friction is proportional to the mass ratio between the host and the satellite halos, meaning that minor mergers (i.e. mergers with small halos) are more affected by this effect. This also causes a delay in the mass buildup of galaxies. Nevertheless, going towards lower redshift and higher masses these differences soon become negligible, since the mass buildup is dominated by high-mass progenitors. The detailed differences between the {\it ins1} and {\it tdf4} scenarios have also been discussed in \cite{dayal2019} and \cite{piana2021}.

\section{Contribution of AGN to the UV luminosity of early galaxies}
\label{UV LF_dis}

We now study the relative contribution of star formation and black holes to the UV luminosities of early galaxies (Sec.~\ref{rel_uv}) before discussing the fractional lifetime of galaxies dominated by UV luminosity from black holes (Sec.~\ref{frac_lt}).

\subsection{Relative contribution of star formation and black holes to the UV luminosity}
\label{rel_uv}

\begin{figure*}
\includegraphics[width=\textwidth]{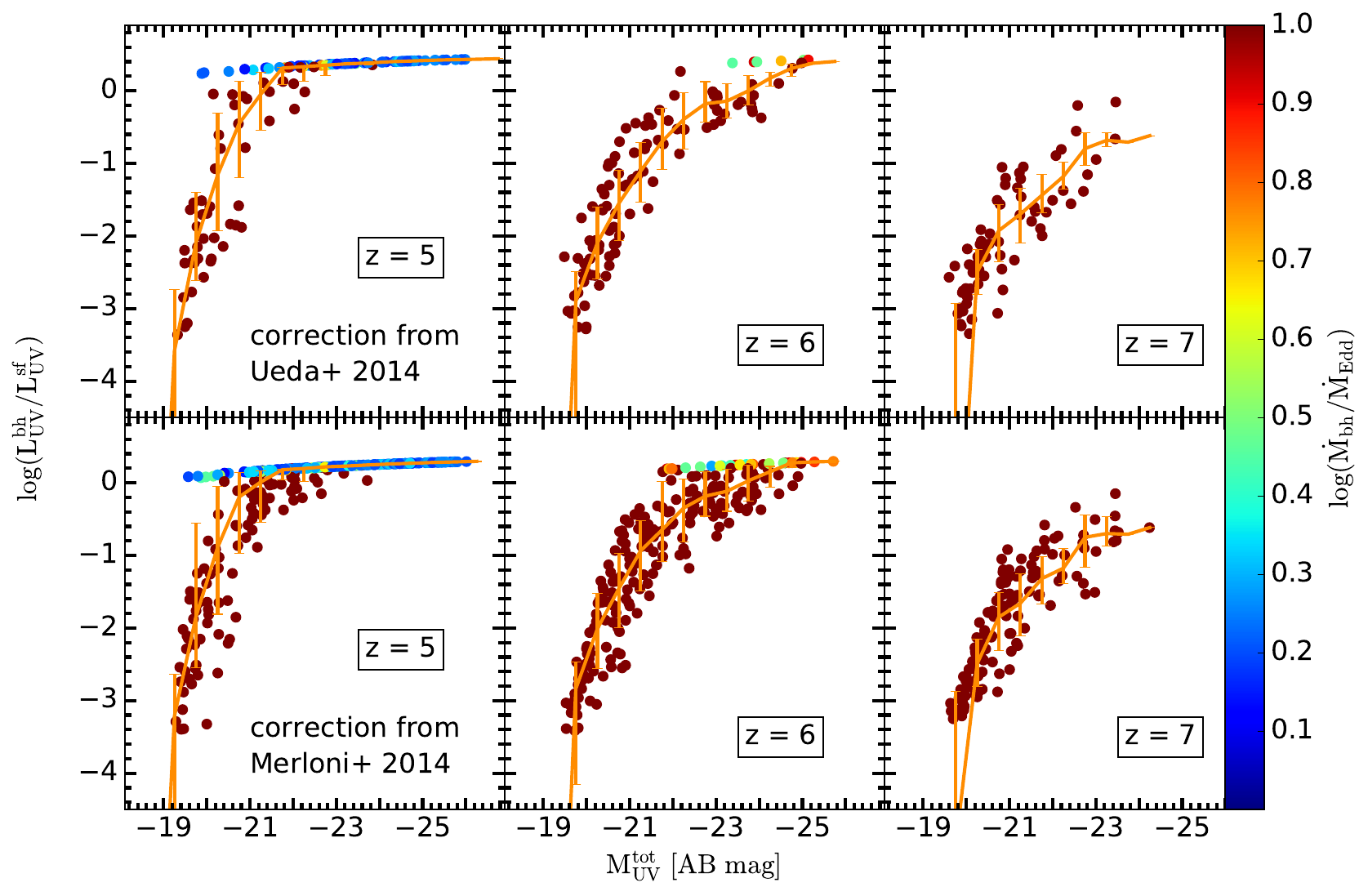}\\
\caption{The ratio between the UV luminosity from black hole accretion and star formation as a function of total UV magnitude for all of the galaxies present at $z = 5-7$ for the black hole correction models noted in each row. We remind the reader that while our star formation efficiency is tuned to reproduce observations (in particular the galaxy UV LF and SMF), we have corrected for AGN obscuration effects by randomly sampling the galaxies in each bin (see Sec.~\ref{sec_model} for more details). The orange solid line represents the median UV ratio averaged over 1000 realisations of the obscuration effects (as explained at the bottom of Sec.~\ref{sec_model}) and the error bars show the average 16th and 84th percentile of the distribution. The data points show instead one random realisation, with the colour coding indicating the black hole accretion Eddington ratio.} 
\label{uvrat}
\end{figure*}

We start by discussing the importance of black hole luminosity in star-forming galaxies. Distinguishing AGN from SF galaxies with the sole use of photometry has proven to be a challenge. This is true especially for galaxies with $M_\mathrm{UV} \gtrsim -23.5$, where LBGs might become dominant \citep{ono2018, adams2020, harikane2021}, given that the AGN UV LF is still very uncertain, with faint-end slopes ranging from $\alpha \sim -1.3$ \citep{matsuoka2018, akiyama2018} to $\alpha \sim -2$ \citep{mcgreer2018, shin2020}. Some studies have tried to impose a condition on the compactness of the source, for it to be considered an AGN \citep{kashikawa2015, matsuoka2018}, but this does not ensure that all and only LBGs are filtered out. For this reason, \cite{bowler2021} opted for a distinction based on the size and morphology of the galaxies, in addition to archival spectral information, to estimate the fraction of AGN per luminosity bin for $M_\mathrm{UV} > -24.5$ at $z=4$. In addition, knowing the AGN or SF-dominated galaxy fraction as a function of luminosity would also help in obtaining better constraints on the bright end of the high-redshift galaxy UV luminosity function by correcting for this possible contaminant \citep[though other sources of uncertainty - e.g. cosmic variance - might still persist, e.g.][]{ono2018}. We now employ our model to predict the fraction of galaxies dominated by stellar activity at $z = 5-7$. It is important to note that our modelled AGN UVLF extends only to $M_\mathrm{UV}^\mathrm{bh} \sim -26$ at $z=5$; we are therefore unable to probe the high-luminosity end of this UV LF.

In Fig.~\ref{uvmag_smass}, we show the {\it intrinsic} UV magnitudes for the contribution from star formation ($M_{\mathrm{UV}}^{\mathrm{sf}}$), black hole accretion ($M_{\mathrm{UV}}^{\mathrm{bh}}$), and their sum ($M_{\mathrm{UV}}^{\mathrm{tot}}$) as a function of stellar mass at $z \sim 5-7$. Starting at $z=5$, we see that the black hole magnitude falls off very quickly at $M_{\mathrm{UV}}^{\mathrm{tot}}\gsim -19$, which corresponds to the halo mass threshold $M_\mathrm{h}^\mathrm{crit}$ below which black holes accrete at $7.5 \times 10^{-5} \dot{M}_\mathrm{Edd}$. Once black holes overcome this threshold and enter the Eddington-limited regime at $M_* \gtrsim 10^9 \msun$, their contribution to the total UV luminosity rapidly grows, spanning a wide UV magnitude range with $M_{\mathrm{UV}}^{\mathrm{bh}} \sim -10~(-22)$ mag for $M_* \sim 10^{9}~(10^{10})\msun$, indicative of their exponential growth. As a consequence, we see a steepening of the $M_{\mathrm{UV}}^{\mathrm{tot}} - M_*$ relation at $M_* \gtrsim 10^9 \msun$: the more massive the host galaxy, the more important the contribution of black hole activity to its intrinsic UV emission. The huge spread in $M_{\mathrm{UV}}^{\mathrm{bh}}$ for $10^9 \msun \lsim M_* \lsim 10^{10} \msun$ corresponds to an increase by up to $2$ magnitudes in the scatter of the $M_{\mathrm{UV}}^\mathrm{sf} - M_{*}$ relation. At the high-mass end of the relation, the AGN contribution grows less steeply but steadily, as black holes enter a slower accretion regime regulated by Eq.~\ref{eq_min}, and eventually the AGN activity starts dominating the UV emission. The qualitative situation is similar at $z \sim 6$ for $M_* \sim 10^{9-10}\msun$ galaxies; the black hole emission outshines the stellar emission by $\approx 1.5$ magnitudes at $M_* \gsim 10^{11} \msun$. Finally, at $z=7$, the black hole emission is sub-dominant with respect to the SF component of the UV magnitude at all masses, as supermassive black holes with $M_\mathrm{bh} \gtrsim 10^6 \msun$ have just started to form at this epoch \citep{piana2021}.   

\begin{figure*}
\includegraphics[width=\textwidth]{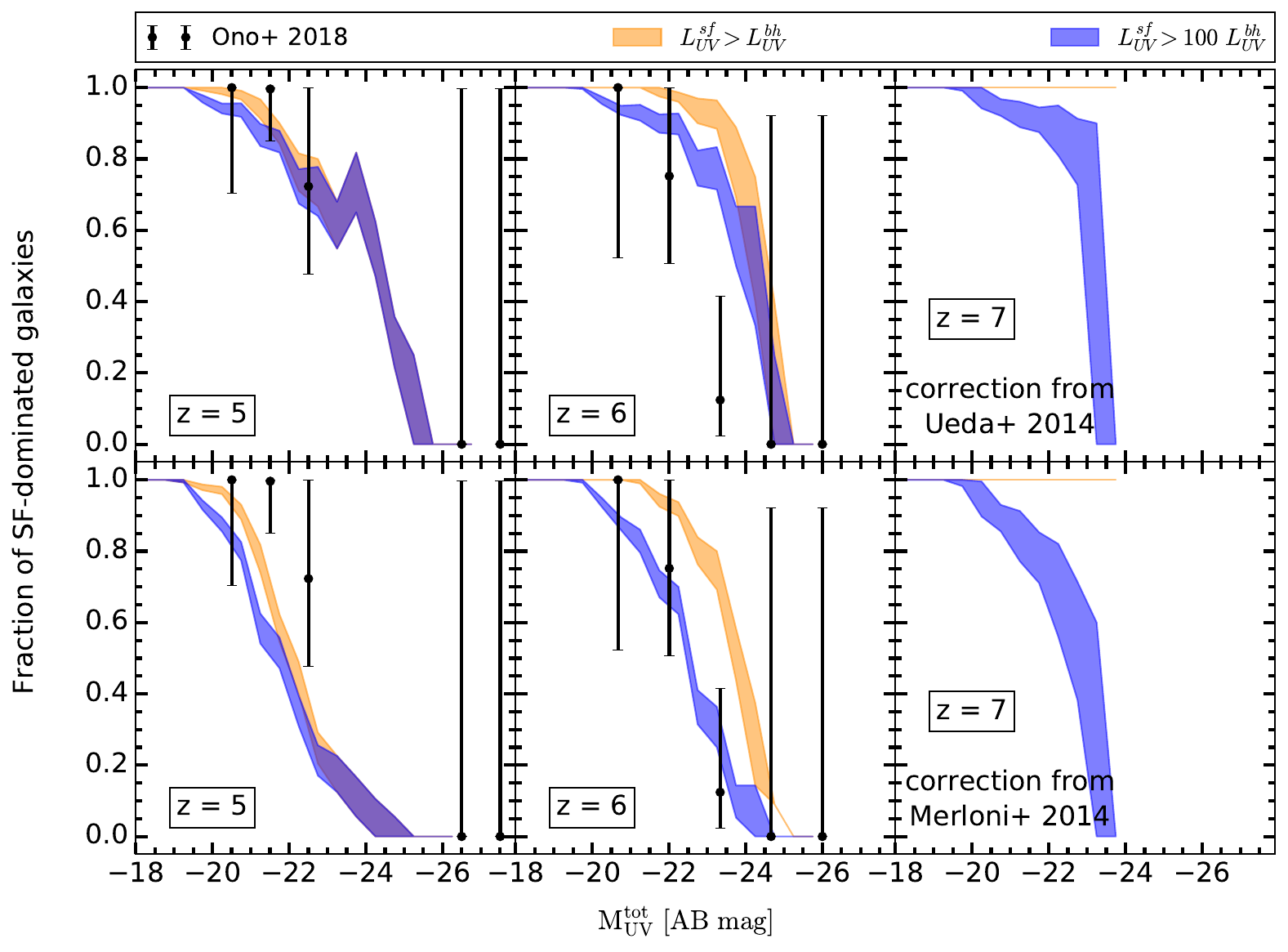}\\
\caption{The fraction of galaxies with a UV ratio $L^\mathrm{sf}_\mathrm{UV} > L^\mathrm{bh}_\mathrm{UV}$ (orange shaded region) and $L^\mathrm{sf}_\mathrm{UV} > 100\ L^\mathrm{bh}_\mathrm{UV}$ (cyan shaded region) as a function of the total UV magnitude at z = 5, 6 and 7; the extent of the shaded areas represents the difference between the 84th and 16th percentile (i.e.~the 1-sigma error) of the distribution in each bin for 1000 realisations of the two black hole obscuration models considered in the upper and lower rows. At $z \sim 5,6$, the solid points show the points collected by \protect\cite{ono2018} with error bars showing the $1-\sigma$ error.}
\label{galfrac}
\end{figure*}

We then study the ratio between the observed UV luminosities of the black hole and stellar components ($L_{\mathrm{UV}}^{\mathrm{bh}} / L_{\mathrm{UV}}^\mathrm{sf}$) as a function of total UV magnitude and Eddington ratio in Fig.~\ref{uvrat} for the AGN obscuration models \citep{ueda2014, merloni2014} discussed in Section~\ref{sec_model}. At all redshifts, we find this ratio to increase with decreasing total UV magnitudes \citep[see also][]{volonteri2017}. Starting with the obscuration fractions from the \citet{ueda2014} model, at $z=5$, Eddington accretion is confined to galaxies with $-23 \lsim M_{\mathrm{UV}}^{\mathrm{tot}} \lsim -19$, where halos have just grown above $M_\mathrm{h}^\mathrm{crit}$. At brighter magnitudes of $M_{\mathrm{UV}}^{\mathrm{tot}} \lsim -23$, galaxies start accreting a fraction ($\sim 10^{-4}$) of the available gas mass instead, as per our formalism. The UV ratio distribution of such galaxies is quite steady with magnitude, since in this regime black holes grow proportional to the gas mass present in the galaxy, like the stellar mass. At $z \sim 6$, most galaxies accrete at the Eddington rate at $M_{\mathrm{UV}}^{\mathrm{tot}} \lsim -20$ (except a few of the brightest ones) with the black hole luminosity becoming increasingly important for brighter galaxies. At $z \sim 7$, while most galaxies continue accreting at the Eddington rate, the black hole luminosity is at most 10\% of the stellar luminosity. Finally, we point out that in the case in which we apply the visibility corrections from \citet{merloni2014} the UV ratio stabilises at a slightly lower value, due to the fact that in order to compensate for a higher fraction of unobscured AGN (see Fig.~\ref{funobs}) we lower the maximum fraction of gas in the host galaxy that the black hole can accrete. In addition, the higher unobscured fraction (see Fig.~\ref{funobs}) leads to more galaxies populating the high-luminosity bins, though the qualitative trends are the same as for the previous case.

In Fig.~\ref{galfrac} we show the fraction of galaxies whose UV luminosity is dominated by the stellar emission, as a function of total UV magnitude. This allows us to perform a comparison with the results obtained by \cite{ono2018}, who measured the fraction of spectroscopically-identified galaxies (intended as the number of spectroscopically-confirmed galaxies divided by that of spectroscopically-confirmed galaxies plus AGN) in high-redshift catalogues, though with wide error bars. Since our model seems to reproduce better these results at $z=5$ and 6 if we define SF-dominated galaxies those for which $L_{\mathrm{UV}}^{\mathrm{sf}} > 100\ L_{\mathrm{UV}}^{\mathrm{bh}}$, we will keep this as reference value also in the next sections to distinguish between SF and AGN-dominated galaxies. In particular, galaxies are fully dominated by SF UV emission for $M_{\mathrm{UV}}^{\mathrm{tot}} \gtrsim -19.5$, while AGN activity dominates the luminosity of all galaxies with $-26 \lsim M_{\mathrm{UV}}^{\mathrm{tot}} \lsim -25$ at $z = 5$ and 6. Also, the magnitude bin in which the numbers of AGN-dominated and SF-dominated galaxies are roughly equal falls around $M_{\mathrm{UV}} \sim -24~(-22)$ both at $z = 6$ and $z=5$ when the Ueda (Merloni) AGN visibility corrections are applied, indicating that there is only minimal redshift evolution of the fraction of SF-dominated galaxies between $z = 6$ and 5. As noted, this result can be useful to estimate the expected fraction of AGN in large high-redshift photometric surveys and to correct the bright end of the galaxy UV LF for this contribution. 

\subsection{Fractional lifetime for AGN domination of UV luminosity}
\label{frac_lt}

\begin{figure}
\includegraphics[width=0.5\textwidth]{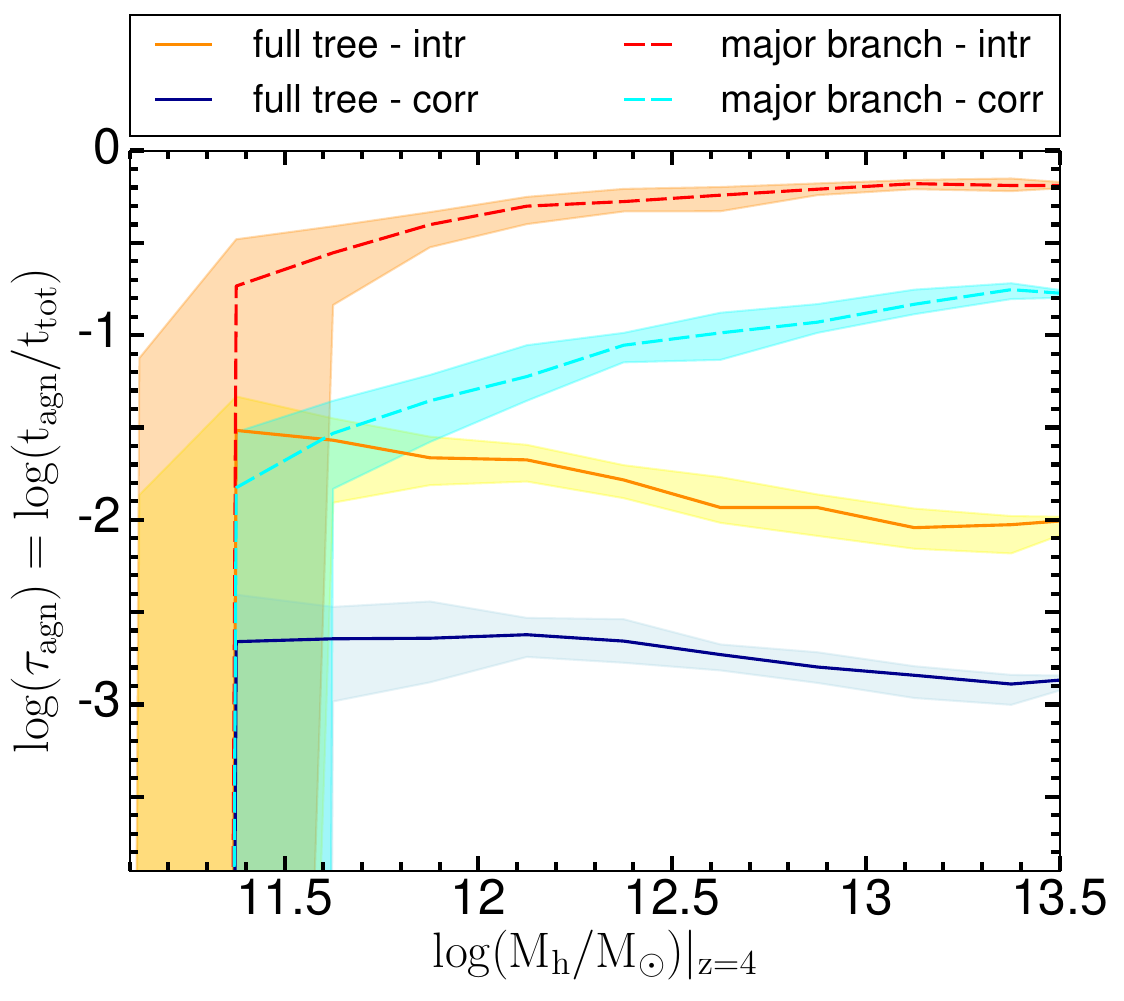}\\
\caption{As a function of the final halo mass at $z=4$, we show the fractional lifetime for which $L_\mathrm{UV}^\mathrm{sf} < 100\ L_\mathrm{UV}^\mathrm{bh}$. The lines represent the median and the shaded areas represent the 16th and 84th percentiles of the distribution within each bin. The dashed and solid lines show results for the major branch and the full tree for the intrinsic and corrected black hole luminosities, as marked. }
\label{agndom}
\end{figure}

We now discuss the fractional lifetime that galaxies spend in the AGN-dominated phase (defined for $L_\mathrm{UV}^\mathrm{sf} \lsim 100\ L_\mathrm{UV}^\mathrm{bh}$), the results for which are shown in Fig.~\ref{agndom}. For each halo at $z=4$ with final mass $M_\mathrm{h} > M_\mathrm{h}^{\mathrm{crit}}$ we walk through the merger tree, computing the fraction of time spent in the AGN-dominated phase by all of its progenitors (solid lines) and just the major branch (dashed lines). Considering the full merger tree and no dust attenuation of the black hole luminosity, galaxies with $M_\mathrm{h} \sim 10^{13.5} \msun$ spend only about $1\%$ of their cumulative lifetime in black hole UV luminosity dominated mode, since their progenitors mass function is mostly dominated by low-mass halos. On the other hand, lower-mass halos with $M_\mathrm{h} \sim 10^{11.5} \msun$ live in this phase three times as long, due to their lower number of progenitors and hence the fact that their total average lifetime is shorter. When visibility corrections are implemented, the difference between the high-mass and low-mass ends is smaller due to the fact that the obscured fraction increases at lower-masses. In general, dust corrections bring down the black hole UV luminosity-dominated fractional lifetime by approximately a factor of ten at all masses, since dust obscuration effects apply only to black hole luminosity and not to the stellar luminosity. Looking at the major branches of the halos, the AGN duty cycle $\tau_\mathrm{agn}$ is $\log(\tau_\mathrm{agn}) \sim 0.64~(0.18)$ for the progenitors of halos with $M_\mathrm{h} \approx 10^{13.5}~(10^{11.5}) \msun$ in the intrinsic case, since the major branch of higher-mass halos enters the Eddington-limited accretion phase at earlier times \citep{piana2021}. When obscuration effects are applied, these numbers go down by a factor of 4~(12) for halos with $M_\mathrm{h} \approx 10^{13.5}~(10^{11.5}) \msun$, as expected by the mass dependence of the obscured fraction of AGN, and in qualitative agreement with the results found in \cite{trebitsch2019}. Finally, the variance of $\tau_\mathrm{agn}$ within each mass bin is directly linked to the specific variations in the mass assembly history (especially for halos with final masses $M_\mathrm{h} \lsim 10^{12.3} \msun$).

\subsection{Build-up of the AGN and SF UV LF}

\begin{figure*}
\includegraphics[width=\textwidth]{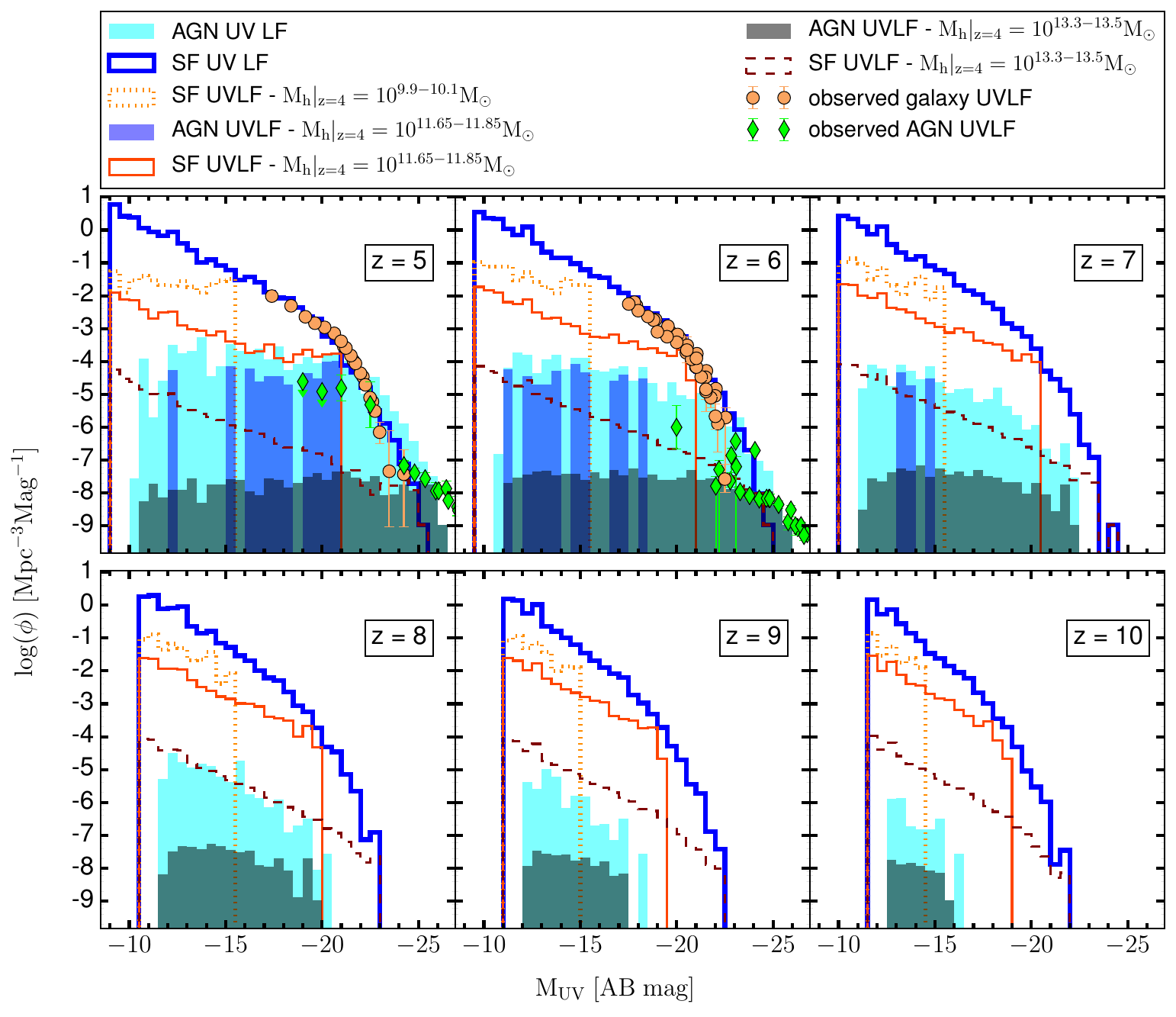}\\
\caption{The evolving UV LF from $z \sim 5-10$, as marked, for black hole accretion and star formation. We show how the progenitors of galaxies with $z=4$ halo masses $M_\mathrm{h} = 10^{9.9-10.1}\msun$ (dotted lines), $M_\mathrm{h} = 10^{11.65-11.85}\msun$ (thin solid lines) and $M_\mathrm{h} = 10^{13.3-13.5}\msun$ (dashed lines) populate the intrinsic star formation (thick solid blue line) and black hole (thick solid cyan) UV LF (blue empty histogram). We also show observational results for the AGN UVLF at $z=5$ \citep{mcgreer2013, parsa2018} and $z=6$ \citep{willott2010, kashikawa2015, jiang2016, parsa2018, matsuoka2018} and for the galaxy UVLF at $z=5$ \citep{bouwens2015, ono2018} and $z=6$ \citep{bouwens2007, mclure2009, willott2013, bouwens2015, finkelstein2015}.}
\label{UV LF_evolution}
\end{figure*}

We now study the evolving AGN and LBG UV LF from $z \sim 5-10$ as shown in Fig.~\ref{UV LF_evolution}. The shape of the high-redshift LBG UV LF has now been extremely well-constrained observationally within the magnitude interval $-22.6 \lsim M_{UV} \lsim -14.5$ \citep{oesch2010, mclure2013, bradley2014, atek2015, bowler2015, mcleod2016, bouwens2017, livermore2017}. However, it remains debated as to whether its evolution is primarily driven by an evolution of the underlying number density \citep{ouchi2009, mclure2009} or of the characteristic knee luminosity \citep{castellano2010, bouwens2011b, grazian2011, bouwens2012b}. On the other hand, the shape of the AGN UV LF remains debated at high-z, ranging from a Schechter function \citep{manti2017} to a double-power law \citep{jiang2016, parsa2018, kulkarni2019}.

In Fig.~\ref{UV LF_evolution} we show the LBG and AGN UV LF at $z \sim 5-10$, deconstructed into the contributions from the progenitors (including all the minor branches as well) of the three halo mass bins discussed in the previous section. Starting with the LBG UV LF, we find that while the progenitors of $M_\mathrm{h}|_{z=4} = 10^{9.9-10.1}\msun$ halos have UV magnitudes $-14.5 \lsim M_\mathrm{UV}^\mathrm{sf} \lsim -11.5$ at $z \sim 10$, these widen to $-15.5 \lsim M_\mathrm{UV}^\mathrm{sf} \lsim -9$ by $z \sim 5$. At all redshifts, they dominate the number density at the faintest end of the UV LF. As expected, the progenitors of $M_\mathrm{h}|_{z=4} = 10^{11.65-11.85}\msun$ halos extend to brighter luminosities, with $M_\mathrm{UV}^\mathrm{sf} \sim -19 (-21)$ at $z \sim 10 (5)$, dominating the UV LF number density in the intermediate UV range. Finally, the progenitors of halos as massive as $M_\mathrm{h}|_{z=4} = 10^{13.3-13.5}\msun$ are responsible for building-up the brightest end of the UV LF. This confirms a picture wherein the evolution of the UV LF depends on the luminosity range probed \citep[see also][]{Dayal2013}: the evolution of the bright end of the LF is due to a fairly steady increase in the UV luminosity in the most massive LBGs. The evolution of the low-luminosity end, on the other hand, involves a mix of positive and negative luminosity evolution (as low-mass galaxies temporarily brighten then fade) coupled with a positive and negative density evolution as new low-mass galaxies form at each redshift while others are consumed by merging.

As for the AGN UV LF, we start by noting that its amplitude exceeds that of the LBG UV LF at $z \sim 5-6$ for $ M_\mathrm{UV} \lsim -23$. However, as a result of the black hole-growth conditions imposed by our model (where black holes in halos below a characterisic mass can only grow at $7.5 \times 10^{-5}$ of the Eddington rate), the LBG UV LF dominates at all magnitudes at $z \gsim 7$. We also find that the progenitors of $M_\mathrm{h}|_{z=4} = 10^{9.9-10.1}\msun$ halos, which spend their lives below the halo mass threshold, do not contribute to the considered AGN UV LF, since their black holes are stuck in the low Eddington regimes and their magnitudes is $M_\mathrm{UV}^\mathrm{bh} \gtrsim -10$. For halos in the intermediate mass bin, as we showed in \cite{piana2021}, only the major branches of these halos grow above $M_\mathrm{h}^\mathrm{crit}$, and they do so at different epochs according to the details of their assembly history. This means that the progenitors of these halos enter the AGN UV LF at different times, and sit in different magnitude bins across the range $-21 \lsim M_\mathrm{UV} \lsim -12$ ($-18.5 \lsim M_\mathrm{UV} \lsim -11$) at $z = 5~(6)$. Interestingly, we find that the AGN UVLF of the progenitors of the highest-mass bin is relatively flat at all epochs, especially for $z \lsim 7$ and $M_\mathrm{UV} \gtrsim -22$. This is because these AGN, accreting at Eddington rate, reside in halos of relatively similar masses (hence number densities: see Fig.~\ref{uvmag_smass}) with $M_\mathrm{h} \gtrsim M_\mathrm{h}^\mathrm{crit}$. On the other hand, the highest-luminosity end of the AGN UVLF is populated by black holes accreting proportional to the gas mass present in the host galaxy, and hence tracks the exponential fall-off of the halo mass function. 

\section{Impact of black holes on the assembly of early galaxies}
\label{bh_impact}

We can now discuss the impact of black hole growth on the gas content (and hence star formation rates) of their hosts, the mass loading factors for SNII and black hole feedback and the impact of black hole feedback on the star formation rate-stellar mass relation.

\subsection{Black hole impact on the gas mass assembly of early galaxies}

\begin{figure*}
\includegraphics[width=0.9\textwidth]{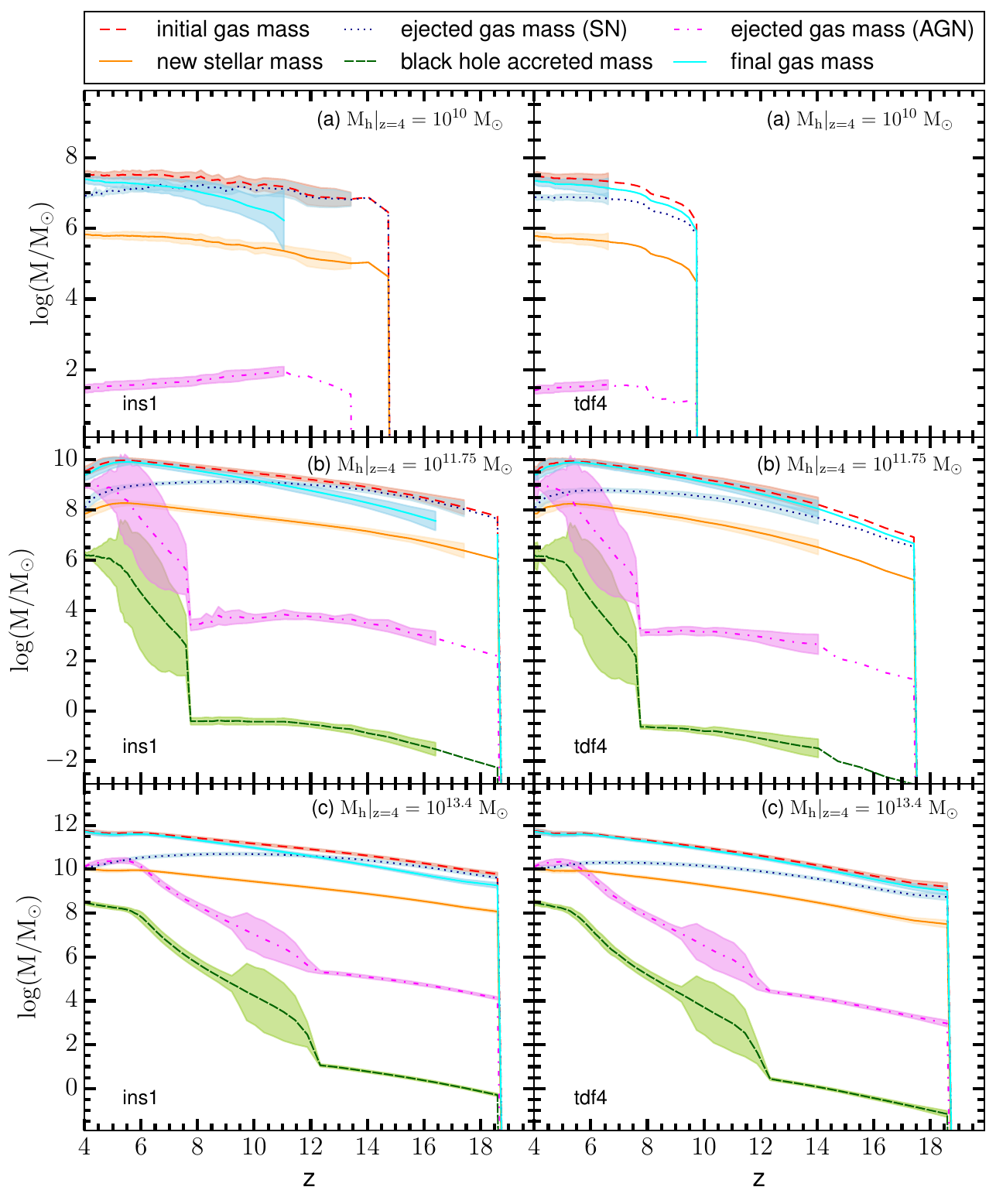}\\
\vspace{-3mm}
\caption{The mass accretion histories for halos of three different halo masses at $z=4$. From top to bottom: $M_\mathrm{h}|_{z=4} = 10^{10}$, $10^{11.75}$, $10^{13.4}\Msun$), each averaged over 21 different halos of similar mass. On the left panels we follow, in the $ins1$ case, the differential evolution of the gas mass present at the beginning of the time step ($M_{\mathrm{gi}}$), the newly-formed stellar mass ($M_{\mathrm{*}}$), the instantaneous gas mass ejected because of SN feedback ($M^\mathrm{ge}_\mathrm{*}$), the mass accreted by the black hole ($M^\mathrm{ac}_\mathrm{bh}$), the gas mass ejected from the galaxy because of black hole feedback ($M^\mathrm{ge}_\mathrm{bh})$ and the gas mass left in the host at the end of the time step ($M_\mathrm{bh}^\mathrm{gf}$). In the right panels we show the same quantities for the $tdf4$ scenario. The shaded areas represent the standard deviation of the averaged sample within each time step. We highlight that $M_{\mathrm{gi}}$ grows with time as galaxies assemble their halos, closely followed by $M_{\mathrm{*}}$. For the intermediate-mass bin, $M^\mathrm{ge}_\mathrm{bh}$ at $z \sim 4$ is high enough to negatively impact the star formation rate, while no significant decrease is noticed for the SFR in the highest-mass bin.}
\label{mah_all}
\end{figure*}

We start by tracking the gas and stellar mass assembly of early galaxies (in the same halo mass bins discussed above) to understand the impact AGN feedback, the results for which are shown in Fig.~\ref{mah_all}. We show results for both the {\it fiducial} model (left column) and the {\it tdf4} model (right column). Starting with the {\it fiducial} model, we see that the initial gas mass ($M_{\mathrm{gi}}$) grows with time as galaxies assemble their halos and gas mass through accretion and mergers. The trend of the newly-formed stellar mass ($M_{\mathrm{*}}$) closely follows that of the initial gas mass at all halo masses, with the star formation efficiency being capped at $2\%$. In turn, the trend of the ejected gas mass due to SN feedback ($M^\mathrm{ge}_\mathrm{*}$) depends on $M_{\mathrm{*}}$. At early times and low masses, $M^\mathrm{ge}_\mathrm{*}$ is of the same order of magnitude as the initial gas mass, and towards $z=4$ it steadily decreases with increasing mass, down to $\sim 30\%$ (upper panels), $3\%$ (middle panels) and $1.5\%$ (bottom panels) of $M_\mathrm{gi}$ respectively for each mass bin. This is easily explicable by the fact that, as the halo potentials deepen, galaxies are able to retain more of their gas mass after SNII feedback. This is linked to the fact that in our model the efficiency with which SN feedback ejects gas mass directly depends on the rotational velocity of the host halo: higher halo masses correspond to higher rotational velocities and hence to higher binding energies \citep[see also][]{dayal2019}.

With respect to black holes, as also discussed above, their growth is fully suppressed for the lowest mass bin where halo masses are less than $M_\mathrm{h}^\mathrm{crit}$. As a result, black holes can only accrete at $7.5 \times 10^{-5}$ of the Eddington rate. Black hole accretion becomes relevant for $M_\mathrm{h}|_{z=4} \gsim 10^{11.75}~(10^{13.5})\msun$ halos at $z \sim 8~(12)$ when the host halos grow above $M_\mathrm{h}^\mathrm{crit}$ and black holes can start accreting either at the Eddington rate or a fraction of the available gas mass. The ejected gas mass due to black hole feedback ($M^\mathrm{ge}_\mathrm{bh}$) is proportional to the gas mass accreted by the black hole $M^\mathrm{ac}_\mathrm{bh}$, and it tends to decrease with decreasing redshift and the growth of the potential well. In particular, for the intermediate (highest) mass bin, black hole feedback starts dominating the instantaneous budget of the expelled mass at $z = 5~(6)$, with the final $M^\mathrm{ge}_\mathrm{bh}/ M^\mathrm{ge}_\mathrm{*}$ ratio reaching $\approx 10~(1.5)$ at $z=4$. In fact, the highest mass bin is characterised by a higher fraction of low-mass progenitors which do not host any black hole activity, thus showing a lower $M^\mathrm{ge}_\mathrm{bh}/ M^\mathrm{ge}_\mathrm{*}$ ratio at $z=4$ with respect to the intermediate-mass bin. 

We briefly touch upon the impact of black hole feedback on the star formation rates - this is discussed in more detail in Section~\ref{sec_ms} that follows. For intermediate-mass halos, we also see that the increasing trend of the gas mass ejected by black hole feedback affects the initial gas mass - and hence the star formation rate - at $z \lsim 5$, with $M^\mathrm{ge}_\mathrm{bh} \approx 0.1~(0.3) \times M_{\mathrm{gi}}$ at $z = 5~(4)$. In the high-mass bin, instead, we see that $M^\mathrm{ge}_\mathrm{bh} \approx 0.1~(0.05) \times M_{\mathrm{gi}}$ at $z = 6~(4)$. The huge variance in the black hole accretion rate (and in the corresponding feedback) represents the fact that black holes residing in halos of similar final mass enter the Eddington-limited accretion phase at different times, depending on the build-up of the host halo mass. We can then define the final gas mass recovered at the end of the time step as $M^\mathrm{gf}_\mathrm{bh} = M_{\mathrm{gi}} - M_{\mathrm{*}} - M^\mathrm{ge}_\mathrm{*} - M^\mathrm{ac}_\mathrm{bh} - M^\mathrm{ge}_\mathrm{bh}$, which yields $M^\mathrm{gf}_\mathrm{bh} \sim 0$ at $z \gtrsim 11~(16)$ for halos with $M_\mathrm{h} \sim 10^{10}~(10^{11.75}) \msun$ and $M^\mathrm{gf}_\mathrm{bh} \sim M_{\mathrm{gi}}$ at $z \lsim 10~(8)$ for halos with $M_\mathrm{h} \sim 10^{11.75}~(10^{13.4})$.

Finally, galaxies show a delay in the mass assembly of halos in the low (intermediate) mass bin for the {\it tdf4} model. As shown in the same figure, low-(intermediate-) mass galaxies start assembling at $z \sim 9.5~(17.5)$ in the {\it tdf4} model as compared to $z \sim 14.5~(18.5)$ in the {\it fiducial} case. Nevertheless, this difference becomes negligible at lower redshifts, when the mass budget is dominated by massive galaxies which are unaffected by reionization feedback and delayed mergers. In fact, the delay induced by dynamical friction increases with the host-to-satellite halo mass ratio, which means only minor mergers with the smallest halos are affected. For this reason, we expect no significant change in the statistical properties of galaxies or AGN at $z \lsim 10$. 


\begin{figure}
\includegraphics[width=0.5\textwidth]{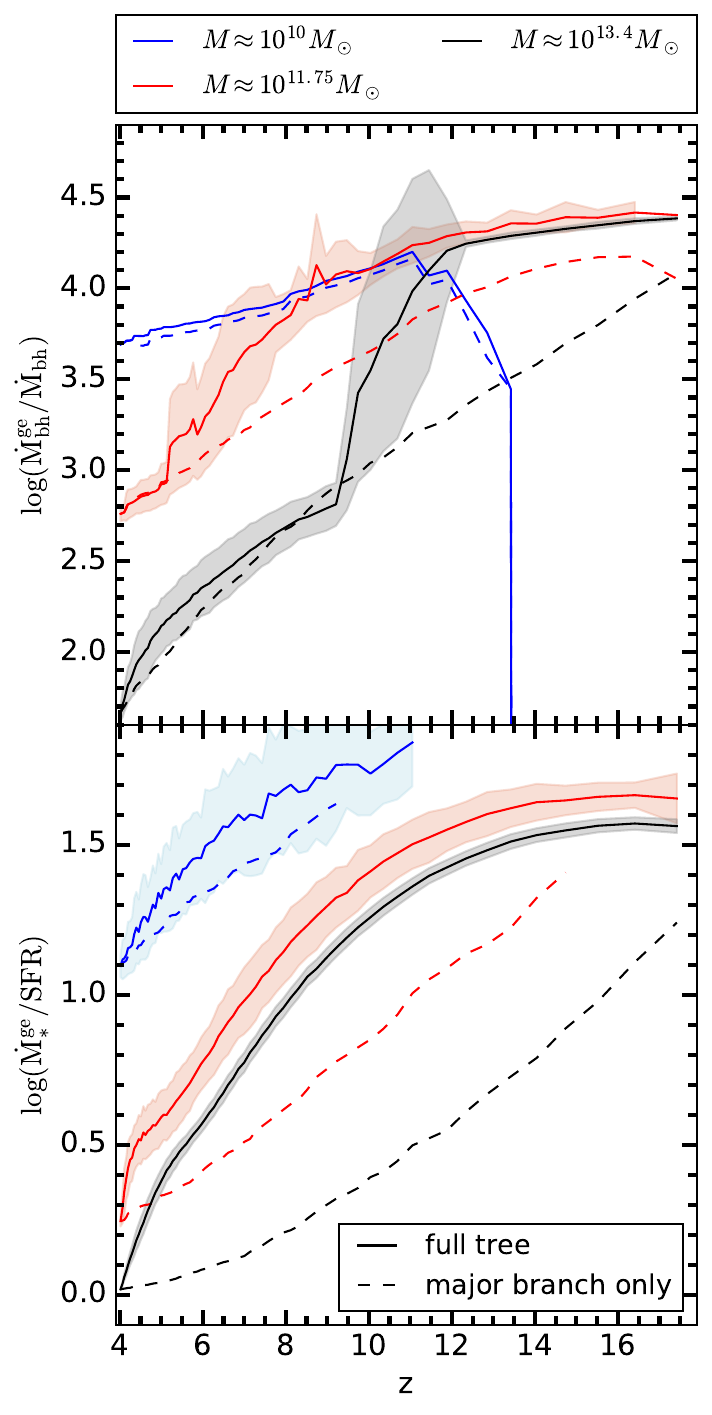}\\
\caption{Time evolution of the mass loading factors both for black holes (upper panel) and star formation (lower panel) averaged over the 21 halos for the three different halo mass ranges considered in Fig.~\ref{mah_all}. The solid lines represent the total mass loading factors obtained summing up - at each redshift - the mass components of all the progenitors of a halo; dashed lines show mass loading factors considering the major branch of the merger tree only.}
\label{masslf2}
\end{figure}

\subsection{Mass loading factors for star formation and black hole feedback}

We now look at the efficiency with which star formation and black hole activity drive gas outflows in Fig.~\ref{masslf2}. We define the mass loading factor for star formation $\overline{\eta}_*(z)=\sum\dot{M}_*^\mathrm{ge}(z)/\sum\mathrm{SFR}(z)$ as the ratio between the gas outflow rate due to SNII feedback and the star formation rate, while that for black holes $\overline{\eta}_\mathrm{bh}(z)=\sum\dot{M}_\mathrm{bh}^\mathrm{ge}(z)/\sum\dot{M}_\mathrm{bh}(z)$ corresponds to the ratio between the ejected and accreted mass by the black hole. The sums run over all the progenitors of a halo at a specific redshift $z$.

When we compute the mass loading factors across the full merger trees, we see that black hole accretion is much more efficient than star formation in driving outflows per unit of mass, and given the same halo mass bin we have that $\eta_\mathrm{bh} \approx 10^{2-3} \times \eta_\mathrm{*}$ at all redshift, despite the higher coupling factor for SN feedback (we remind the reader that we set $f_*^\mathrm{w} = 0.1$ and $f_\mathrm{bh}^\mathrm{w} = 0.003$). This is expected since the efficiency with which mass is turned into energy is much greater for black holes than for Supernovae. In both cases, the mass loading factor is defined by the balance between the energy emitted by the growing black hole (stellar) mass on one hand, and the continuously increasing binding energy of the host halo on the other. As a consequence, both the SN and AGN mass loading factors decrease towards lower redshift and for higher mass bins.

In more details, the lowest-mass bin shows a delayed formation of black holes, as we have seen in Fig.~\ref{mah_all}. At $z>11$ these black holes do not have gas to accrete, since they reside in halos whose gas mass has been fully ejected by SN feedback. At $z \sim 11$, halos can retain some of the gas after star formation but they are still below $M_\mathrm{h}^\mathrm{crit}$, and $\eta_\mathrm{bh} \sim 10^4$. From this point onward the trend steadily decreases with the increasing host mass, down to $10^{3.7}$ at $z=4$. In the intermediate- (high-) mass bin, $\eta_\mathrm{bh}$ also shows a decreasing trend. At $z \sim 8~(12)$ for intermediate- (high-) mass halos we have an exponential increase in $\dot{M}_\mathrm{bh}$ due the start of the Eddington-limited accretion phase, and the steepening becomes ever more pronounced, dropping below $10^4$. The break visible at $z = 5~(9)$ at values of $\eta_\mathrm{bh} \sim 10^3~(10^{2.8})$ for the intermediate- (high-) mass bin is due to the major branches of the trees entering the self-regulated regime and growing less efficiently (i.e.~with lower Eddington ratios) proportionally to the gas mass. As the potential well of the host halo continues to deepen, especially for the high-mass bin, the decreasing trend steepens again. At $z=4$ the difference between the intermediate- and high-mass bin is of more than one order of magnitude, with $\eta_\mathrm{bh} \sim 10^{2.75}$ and $\eta_\mathrm{bh} \sim 10^{1.5}$ respectively.

With respect to the SN mass loading factor, instead, the difference between the intermediate- and high-mass bins is only $\approx 0.2$ orders of magnitude, constant across redshift, and at $z=4$ we have that $\eta_* \sim 1.8$ and $\eta_* \sim 1$ respectively. In fact, the star formation activity, unlike black hole activity, is dominated by the contribution from low-mass progenitors. For the same reason, the major branch plays a lesser role in determining the trend of the SN mass loading factor less than for the black hole mass loading factor. The lowest-mass bin consistently shows a higher value, with $\eta_* \sim 13$ at $z=4$, since they have shallower potential wells.

If we take into account only the major branch of the different merger trees, the evolution of both the AGN and SN mass loading factor is straightforward, and we can write, at each redshift,
\begin{equation}
\eta_*(z) = \frac{\dot{M}_*^\mathrm{ge}(z)}{SFR(z)} = \frac{1 - f_*^\mathrm{eff}}{f_*^\mathrm{ej}(z)} \sim \frac{v_\mathrm{c}^2(z) + f_*^\mathrm{w} v_\mathrm{s}^2}{v_\mathrm{c}^2(z)},
\end{equation}
where $f_*^\mathrm{w} v_\mathrm{s}^2$ is the SN energy density (i.e.~per unit of formed stellar mass) that couples to the gas and $v_\mathrm{s}$ is the velocity of the SN-driven winds, taken to be $611\ \mathrm{km\ s^{-1}}$. Similarly, for the AGN mass loading factor limited to the evolution of the major branch we have
\begin{equation}
\eta_\mathrm{bh}(z) = \frac{\dot{M}_\mathrm{bh}^\mathrm{ge}(z)}{\dot{M}_\mathrm{bh}(z)} = \frac{v_\mathrm{c}^2(z) + \epsilon_\mathrm{r} f_\mathrm{bh}^\mathrm{w} c^2}{v_\mathrm{c}^2(z)},
\end{equation}
where $\epsilon_\mathrm{r} f_\mathrm{bh}^\mathrm{w} c^2$ represents the black hole energy density coupling to the gas per unit of accreted mass. We can then see how in both panels the mass loading factor relative to the major branches decreases with increasing rotational velocity of the halo.

\begin{figure*}
\includegraphics[width=\textwidth]{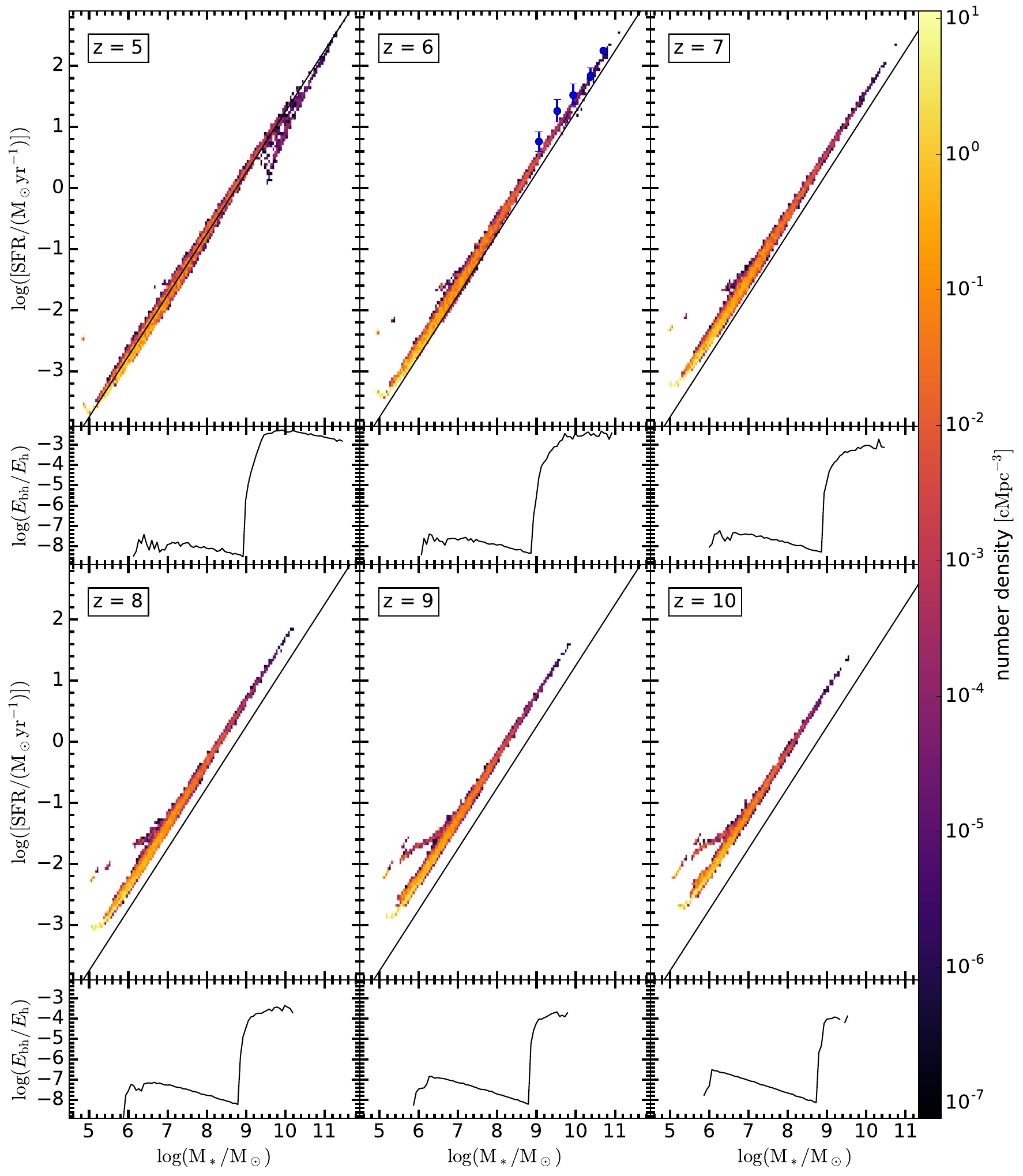}\\
\caption{Density plot of the $M_*$-SFR relation at $z = 5-10$, as marked. Underneath each panel we plot the logarithm of the average black hole accretion energy-to-halo binding energy ratio as a function of stellar mass. The solid black line represents our relation at $z=5$ and follows the relation  $\log(\mathrm{SFR}) = \log(M_*/\msun) - 8.75$, while the blue dots are observational data taken at $5.6 < z < 6.6$ \citep{khusanova2020}.}
\label{gal_ms}
\end{figure*}

\subsection{Impact of black hole feedback on the $\mathrm{M_*-SFR}$ relation}
\label{sec_ms}

To better visualise the impact of black hole feedback on star formation, we show in Fig.~\ref{gal_ms} the distribution of our galaxies in the stellar mass-star formation rate plane in the redshift range $z=5-10$; the colour coding represents the logarithm of the number density of galaxies in each bin. As expected, low-mass galaxies dominate the distribution in terms of number density. In addition, the low-mass end shows a double tail at $z \sim 8-10$, due to the fact that low-mass galaxies, after a first star formation burst, might eject most of their gas and quench their SF activity. At $z\sim 5$, galaxies with $M_* \sim 10^{5-9}\msun$ follow the relation $\log(\mathrm{SFR}) = \log(M_*/\msun) - 8.75$. The $y$-intercept of this relation increases with increasing redshift, and at $z=10$, the star formation rates are a factor of 3 higher than at $z=5$ for a given stellar mass. Galaxies with $M_* \gsim 10^{9}\msun$ at $z \sim 5$ exhibit a decrease in the star formation rate compared to this relation. This is because these galaxies are hosts to active black hole which eject considerable amounts of gas mass. We show this by plotting, underneath each panel, the logarithm of the average ratio between the black hole feedback energy coupling to the gas and the host halo binding energy, as a function of stellar mass. Very low-mass galaxies with $M_* \lsim 10^6 \msun$ host black holes which are not accreting at all and whose growth is fully suppressed, due to the fact that the gas in these galaxies has been completely ejected by SN feedback. When the energy released by the accreting black hole reaches $\approx 10^{-2.5}$ times the binding energy of the gas the host halo, at $z \sim 6$, the galaxy starts straying from the sequence. At $M_* \gtrsim 10^{10.5} \msun$, on the other hand, the binding energy of the halos is high enough to retain the gas within their potential well, the $E_\mathrm{bh}/E_\mathrm{h}$ ratio decreases down to $\sim 10^{-3}$, and the star formation rates can grow up to $\sim 10^3 \msun\ \mathrm{yr^{-1}}$. Our results suggest that AGN feedback, at least at $z>6$, is more effective for the intermediate-mass range, rather than for the high-mass range. In fact, the Eddington ratio (i.e.\ the ratio between the mass accretion rate of the black hole and the Eddington rate) peaks at unity for $M_* \sim 10^{9.5} \msun$, and decreases going towards higher stellar masses \citep{piana2021}, while the coupling factor between the energy emitted by the accreting black hole and the gas remains the same. This seems to imply that the quenching of the highest-mass galaxies occurs either at lower redshift or through different mechanisms. Finally, we point out that our results at $z=6$ are in good agreement with the observed $M_*$-SFR relation inferred in \cite{khusanova2020}.

\section{Conclusions and discussion}

In this work we use the \textit{Delphi} semi-analytic model of galaxy formation and evolution to address the impact of black hole activity on the host galaxies at $z > 4$, following the mass and luminosity build-up of the AGN and stellar components across cosmic time. In order to allow a comparison with observation, we implement the AGN visibility corrections taken from \cite{ueda2014} and \cite{merloni2014}. In the first half of the paper, we discuss the contribution of the AGN emission to the total UV luminosity of the galaxy, as well as the build-up of the galaxy and AGN UV luminosity functions, while in the second part we show the impact that AGN feedback has on the mass assembly history of the galaxy. We list here our key results.

\begin{itemize}
\item We find that the observed fraction of galaxies for which the AGN UV luminosity outshines the stellar component by a factor of 100 is dominant for magnitudes $M_\mathrm{UV}^\mathrm{tot} \lsim -24~(-23)$ at $z = 6$ when we employ the Ueda's (Merloni's) correction model, and that this magnitude becomes fainter with time. Our findings are in good agreement with the galaxy fraction observationally inferred in \cite{ono2018}, though the uncertainties on the obscured fraction of AGN, which is sensitive on the sample of high-redshift AGN that has been used to estimate it, prevent us from reaching unambiguous results.

\item The fractional lifetime that the major branches of our merger trees spend in the AGN mode is $\approx 63\%$ for $M_\mathrm{h}|_{z=4} \sim 10^{13.5} \msun$, decreasing with decreasing halo mass down to $\approx 18\%$ for halos with $M_\mathrm{h}|_{z=4} \sim 10^{11.75} \msun$, though the observed duty cycle, after obscuration effects, is respectively of $\approx 16\%$ and $1\%$.

\item We study the build-up of the AGN and stellar UV luminosity functions, finding that the stellar UV LF of the progenitors of low-mass galaxies (driving the evolution of the faint end of the total UVLF) evolves towards brighter luminosities only down to $z \sim 8$, while that of high-mass galaxies (populating the bright end of the total UVLF) does not stop evolving across the redshift range considered. This is an indication of how the galaxy UVLF evolution is driven by actual brightness evolution at the high-luminosity end. An evolution towards fainter luminosities, due to the action of SN feedback in low-mass galaxies, is also found. 

\item According to our model, SN-driven outflows dominate the instantaneous gas outflow rate down to $z \sim 5~(6)$ for galaxies in halos with $M_\mathrm{h}|_{z=4} \sim 10^{11.4}~(10^{13.5}) \msun$. After then, AGN-driven mass outflows take over.

\item We compute the mass loading factor for SN- and AGN- driven outflows, finding higher values for lower-mass galaxies in both cases, with $\eta_\mathrm{bh} \sim 10^{3.7}, 10^{2.8}$ and $10^{1.5}$ respectively for the low, intermediate and high mass bin at $z=5$. Similarly, for SN feedback, we find respectively $\eta_\mathrm{sn} \sim 13, 1.8$ and $1$.

\item We find that the impact of AGN feedback is greater for galaxies in intermediate-mass halos with $M_\mathrm{h}|_{z=4} \sim 10^{11.75} \msun$ than at higher masses: a deeper potential well and hence a higher halo binding energy can retain part of the gas in outflows. In particular, when the energy emitted by the accreting black hole is approximately $1\%$ of the halo binding energy, we see a corresponding decrease in the star formation rate by a factor of 3. This occurs in galaxies of mass $\sim 10^{9.5} \msun$ at $z=5$.
\end{itemize}

To conclude, the reader should keep in mind a few caveats arising from some of the approximations in our model. First, we are not accounting for gravitational recoil kicks and black hole ejection from the host halo during mergers, which could slow down the subsequent growth of the central black hole \citep{blecha2016, izquierdo-villalba2020}. At the same time, we do not model the black hole binary inspiralling phase after the galaxies have merged, nor the final parsec problem, during which stellar scattering processes are needed for black holes to coalesce together. Taking into account these effects, as well as the stochasticity introduced in black hole orbits by the interaction with gas and stars \citep{pfister2019}, would also delay the build-up of black hole mass. We should also mention that we are ignoring the local environment when computing the LW background irradiating our galaxies, which would enhance DCBH seed formation in more biased regions with respect to less biased ones. At the same time, all the first leaves at $z>13$ that do not host a DCBH are populated with a SBH seed. This is motivated by the fact that we assume all the gas mass of these first halos to only have assembled by smooth accretion and the average IGM metallicity is $Z \lsim 10^{-5} Z_\odot$ at $z>13$ \citep[see for instance Fig.\ 9 of][]{ucci2021}. Tracking the metal enrichment of such halos would require simulations capable of following the patchy metal enrichment of the IGM through cosmic time. Given the complexity of such calculations, which are out of the scope of this paper, we defer these to a future work. 

For all the reasons mentioned above, the black hole masses and luminosities that our model yields should be considered as upper limits. It is important to note, though, that these caveats mostly affect the low-mass end of the galaxy and black hole population, while at high masses and lower redshifts black hole growth is dominated by gas accretion, and mergers become less relevant.

\section*{Acknowledgments} 
The authors thank Marta Volonteri for her insightful physical contributions that have added enormously to the quality of the paper and Maxime Trebitsch for his engaging comments. OP and PD acknowledge support from the European Commission's and University of Groningen's CO-FUND Rosalind Franklin program. PD also acknowledges support from the European Research Council's starting grant ERC StG-717001 (``DELPHI") and from the NWO grant 016.VIDI.189.162 (``ODIN"). TRC acknowledges support of the Department of Atomic Energy, Government of India, under project no. 12-R\&D-TFR-5.02-0700.

\section*{Data Availability}
The data underlying this paper, which have been produced by our model, are available on reasonable request to the corresponding author.

\bibliographystyle{mn2e}
\bibliography{lib}

\label{lastpage} 
\end{document}